\documentclass{emulateapj}

\shorttitle{Detectability of Glycine in Solar-type System Precursors}
\shortauthors{Jim\'enez-Serra et al.}

\begin{document}

\title{Detectability of Glycine in Solar-type System Precursors}

\author{Izaskun Jim\'{e}nez-Serra\altaffilmark{1}, Leonardo testi\altaffilmark{1}, Paola Caselli\altaffilmark{2,3} and Serena Viti\altaffilmark{4}}

\altaffiltext{1}{European Southern Observatory, Karl-Schwarzschild-Str. 2, 85748 Garching (Germany); ijimenez@eso.org, ltesti@eso.org}

\altaffiltext{2}{School of Physics \& Astronomy, University of Leeds, LS2 9JT, Leeds (UK)}

\altaffiltext{3}{Max-Planck-Institut f\"ur extraterrestrische Physik (MPE), Gie{\ss}enbachstr., 85741 Garching (Germany); caselli@mpe.mpg.de}

\altaffiltext{4}{Department of Physics \& Astronomy, University College London, Gower Place, WC1E 6BT, London
(UK); sv@star.ucl.ac.uk}

\begin{abstract}

Glycine (NH$_2$CH$_2$COOH) is the simplest amino acid relevant for life. Its detection in the interstellar medium is key to understand the formation mechanisms of pre-biotic molecules and their subsequent delivery onto planetary systems. Glycine has extensively been searched for toward hot molecular cores, although these studies did not yield any firm detection. In contrast to hot cores, low-mass star forming regions, and in particular their earliest stages represented by cold pre-stellar cores, may be better suited for the detection of glycine as well as more relevant for the study of pre-biotic chemistry in young Solar System analogs. We present 1D spherically symmetric radiative transfer calculations of the glycine emission expected to arise from the low-mass pre-stellar core L1544. Water vapour has recently been reported toward this core, indicating that a small fraction of the grain mantles in L1544 ($\sim$0.5\%) has been injected into the gas phase. Assuming that glycine is photo-desorbed together with water in L1544, and considering a solid abundance of glycine on ices of $\sim$10$^{-4}$ with respect to water, our calculations reveal that several glycine lines between 67$\,$GHz and 80$\,$GHz have peak intensities larger than 10$\,$mK. These results show for the first time that glycine could reach detectable levels in cold objects such as L1544. This opens up the possibility to detect glycine, and other pre-biotic species, at the coldest and earliest stages in the formation of Solar-type systems with near-future instrumentation such as the Band 2 receivers of ALMA. 

\end{abstract}


\section{Introduction}
\label{intro}

Glycine (NH$_2$CH$_2$COOH) is the simplest amino acid and a key constituent of living organisms. It is believed that its formation may have occurred in the interstellar medium (ISM) since glycine and other amino acids have been found in meteorites \citep[][]{piz91,ehr01,gla06} and comets \citep[Wild 2;][]{els09}. Laboratory experiments have also reported the formation of amino acids, including glycine, via ultraviolet (UV) and ion photolysis of interstellar ice analogues \citep[][]{mun02,holt05}, supporting the idea that amino acids may have an interstellar origin \citep[][]{ehr00}. The detection of glycine is thus key to understand the formation of pre-biotic molecules in the ISM and their subsequent delivery onto planetary systems. 

Over a decade glycine has extensively been searched for in high-mass star forming regions such as the hot molecular cores in SgrB2, Orion KL and W51 e1/e2 \citep[][]{kuan03,bell08,bell13}. Hot cores, with temperatures $\sim$100-200$\,$K, show a rich chemistry in complex organic molecules (COMs) due to the evaporation of ices from dust grains \citep[][]{van98}. However, although glycine is expected to be released into the gas phase together with other COMs \citep{garr13}, none of the studies performed to detect glycine in hot cores yielded any firm detection \citep[][]{sny05,cun07,jon07}. 

Several challenges are faced in the search of glycine in hot cores. First, the spectral line density in these objects is high, leading to high levels of line blending and line confusion. Second, the linewidths of the molecular line emission in hot cores are broad (several km$\,$s$^{-1}$), which prevents clear identifications of weak lines from less abundant species. Finally, hot cores are typically located at distances $\geq$1$\,$kpc \citep{gar99} and beam dilution limits the detection of low-abundance, large COMs such as glycine. 

\section{Glycine in low-mass star forming regions}
\label{low-mass}

In contrast with their high-mass counterparts, low-mass star forming regions may be better suited for the detection of glycine for several reasons. The level of line confusion, especially at the earliest stages represented by pre-stellar cores \citep[][]{cas02a,cra05}, is low because the measured gas temperatures are $\leq$10$\,$K \citep[][]{cra07,pag07} and the number of molecular lines excited at these temperatures is smaller than in hot cores. The molecular emission in dark cloud cores also shows linewidths $\leq$0.5$\,$km$\,$s$^{-1}$ \citep[e.g.][]{cas02b,cra07}, which allows accurate identifications of the observed transitions since they suffer less from line blending. COMs such as propylene (CH$_2$CHCH$_3$), acetaldehyde (CH$_3$CHO), dimethyl ether (CH$_3$OCH$_3$) or methyl formate (HCOOCH$_3$) have indeed been detected in dark cloud cores such as TMC-1 and B1 \citep[][]{mar07} and the pre-stellar core L1689B \citep[]{bac12}, unexpectedly revealing a high chemical complexity in the cold gas of these objects.

Previous studies toward hot cores mainly targeted glycine lines with frequencies $\geq$130$\,$GHz because their Einstein $A$ coefficients, $A_{\rm ul}$, are $A_{\rm ul}$$\geq$10$^{-5}$$\,$s$^{-1}$, and their upper level energies are E$_{\rm u}$$\geq$60$\,$K, in agreement with the temperatures in these objects \citep[][]{sny05}. However, for pre-stellar cores, the detection of glycine should be attempted via observations of glycine lines with low values of E$_{\rm u}$ while having {\it not too small} $A_{\rm ul}$ coefficients. In the frequency range between 60$\,$GHz and 130$\,$GHz, glycine (conformer I)\footnote{We only refer to glycine conformer I because the ground vibrational level of glycine conformer II lies 700 cm$^{-1}$ ($\sim$1000 K) above that of conformer I.} has several transitions whose energy levels lie below 30$\,$K and whose $A_{\rm ul}$ coefficients are $\geq$10$^{-6}$ s$^{-1}$, only a factor of 10 lower than those of the glycine lines observed in hot cores \citep[][]{kuan03,sny05}. In this Letter, we test the detectability of glycine toward a well-known pre-stellar core, L1544 \citep{cas99}, by performing simple radiative transfer calculations of glycine between 60$\,$GHz and 250$\,$GHz, and by making reasonable assumptions of the dominant chemical processes in pre-stellar cores and of the abundance of glycine on ices and in the gas phase. 

\section{Radiative transfer modelling of glycine in the L1544 pre-stellar core}
\label{pre-stellar}

\subsection{The Model}
\label{mod}

\begin{figure}
\begin{center}
\includegraphics[angle=0,width=0.5\textwidth]{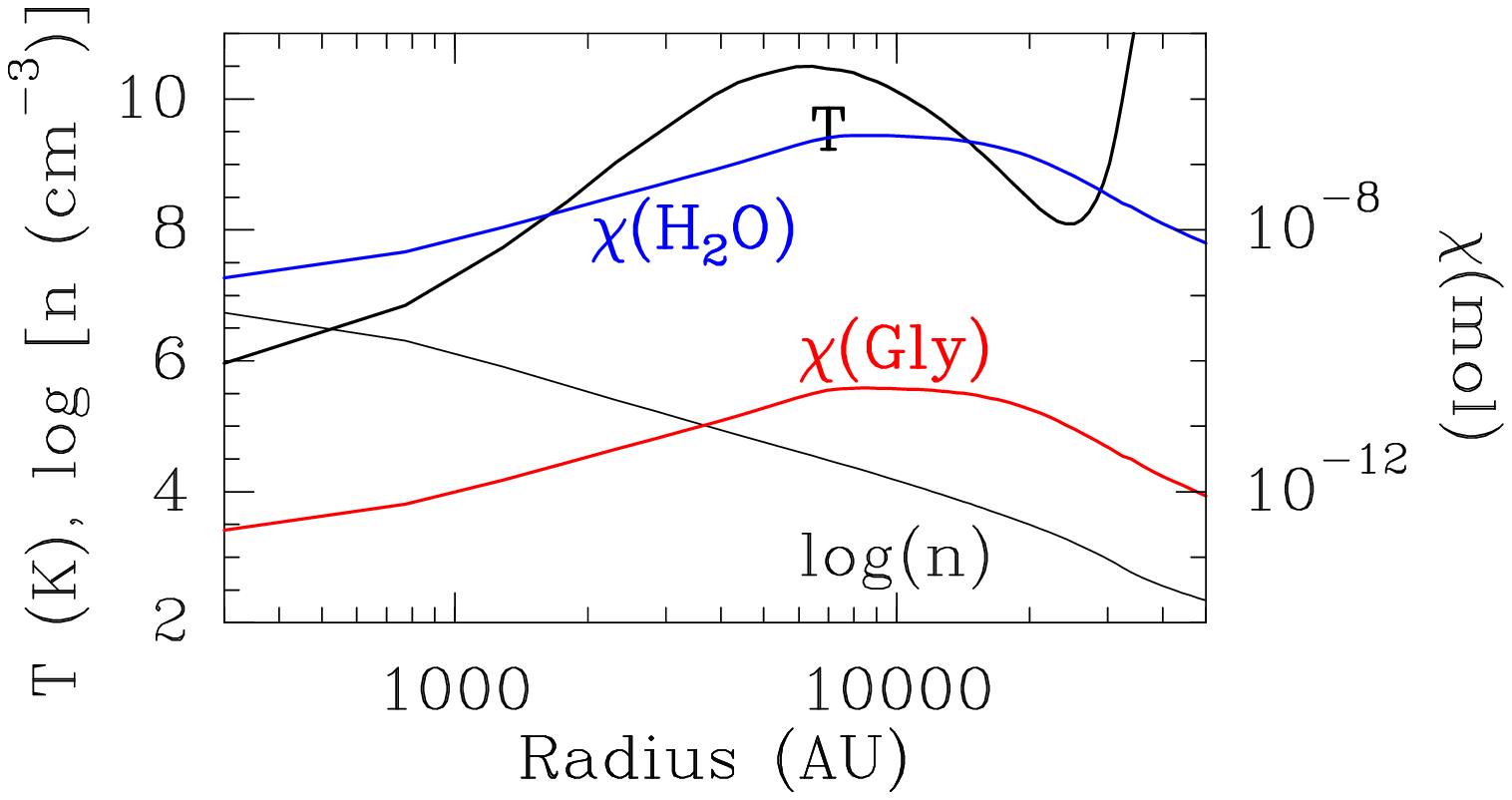}
\caption{1D spherically symmetric distribution of the H$_2$ number density, $n$, and gas temperature, $T$, assumed in our calculations for the L1544 pre-stellar core \citep[][]{ket10}. The radial distribution of water vapour, $\chi$(H$_2$O), was inferred by \citet{cas12} from Herschel HIFI data (blue line). The glycine abundance, $\chi$(Gly), is assumed to follow the distribution of water vapour in L1544, but scaled down by the fraction of solid glycine expected to be formed on ices (red line; see Sections$\,$\ref{mod} and \ref{gly}).}
\label{f1}
\end{center}
\end{figure}

For our calculations, we have considered the L1544 pre-stellar core located in the Taurus molecular cloud \citep[distance of 140$\,$pc;][]{elias78}. This core has largely been studied in the past and its internal physical structure is relatively well-constrained \citep[][]{ward99,cas02a,cas02b,ket10}. Figure$\,$\ref{f1} reports the 1D spherically symmetric distribution of the H$_2$ density, $n$, and gas temperature, $T$, derived by \citet{ket10} for this core, and considered in our calculations. We have also included in the model the gas velocity profile (due to the subsonic phase of the core contraction) and the linewidth of the emission deduced by \citet[][]{cas12}. 

L1544 represents an excellent candidate for our analysis of the detectability of glycine in pre-stellar cores because water vapour has recently been found toward the central few thousand AU of this core \citep{cas12}. 
The distribution of the gas-phase water abundance, $\chi$(H$_2$O), is shown in Figure$\,$\ref{f1} (blue line), where $\chi$(H$_2$O) increases by a factor of $\geq$100 within the central 10000$\,$AU (from $\sim$2$\times$10$^{-9}$ to $\sim$3$\times$10$^{-7}$) to decrease again to $\sim$10$^{-8}$. As proposed by \citet{cas12}, water vapour in L1544 (with $\sim$0.5\% of the total water abundance on ices injected into the gas phase) is produced by the internal FUV field induced by the impact of cosmic rays with H$_2$ molecules. The models of \citet{cas12} do not include the gas-phase chemistry of water. However, their simple calculations not only recover the results from the comprehensive gas-grain models of \citet{hol09}, but also reproduce the inverse P-Cygni profile observed for water in this core. Since gas-phase chemistry is not the dominant process in the production of water vapour in L1544 \citep{cas12,ket14}, we assume that it is mostly produced by comic ray induced FUV photons. We note that this is consistent with the depletion of CO in L1544 \citep[by a factor of 100;][]{cas99} since a higher CO desorption rate \citep[30 times higher than that used by][]{has93} is needed to explain the observed C$^{18}$O and C$^{17}$O lines \citep{ket10}. This indicates that CO is still present in the gas phase in the core center as a result of the ice photodesorption by cosmic ray induced FUV photons \citep{ket10}.

\begin{figure*}
\begin{center}
\includegraphics[angle=0,width=1.0\textwidth]{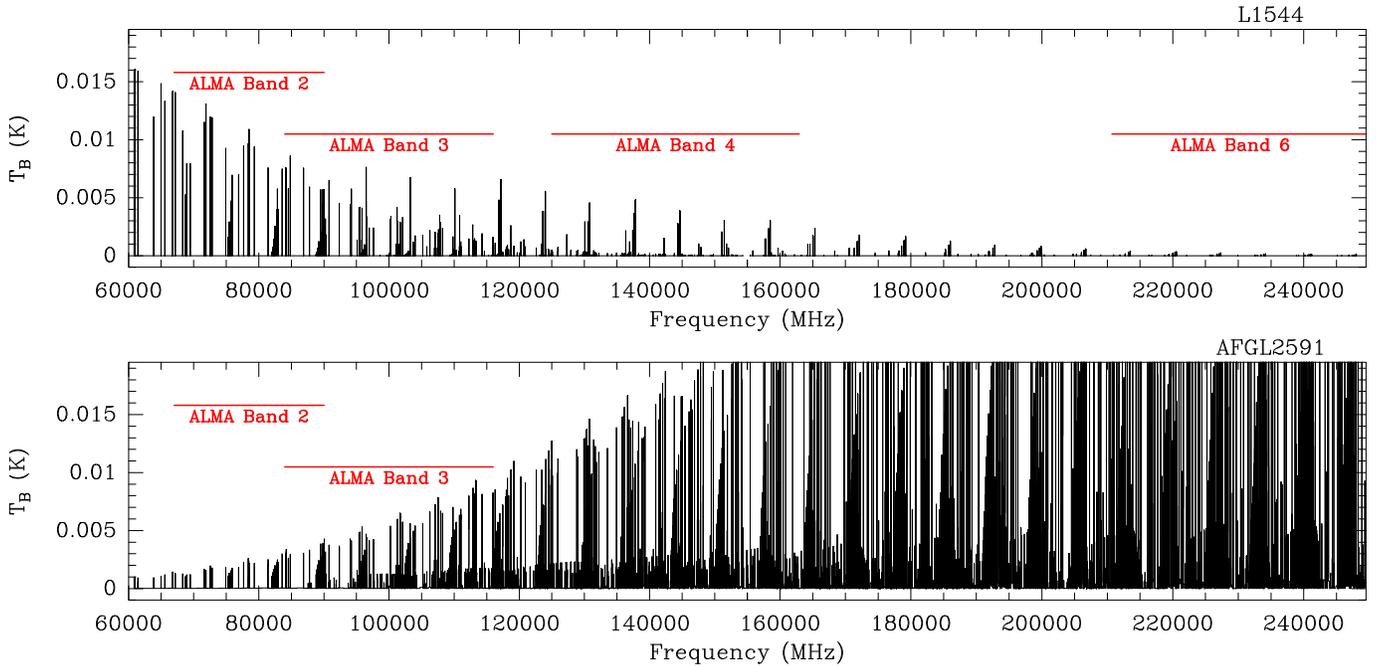}
\caption{{\it Upper panel:} Simulations of the spectrum of glycine (conformer I) obtained for the frequency range between 60$\,$GHz and 250$\,$GHz, considering the physical structure of the L1544 pre-stellar core \citep[Figure$\,$\ref{f1} and][]{cas12}, a solid glycine abundance of $\sim$10$^{-8}$ on ices (Sections$\,$\ref{mod} and \ref{gly}), and LTE conditions. Horizontal red lines indicate the frequency coverage of ALMA Bands 2, 3, 4 and 6. The glycine lines with frequencies 60-80$\,$GHz show peak intensities $\geq$8-10$\,$mK, detectable with near-future instrumentation such as the Band 2 receivers of ALMA. {\it Lower panel:} LTE spectrum of glycine predicted for the same frequency range (60$\,$GHz-250$\,$GHz) but for the AFGL2591 hot molecular core. The hot core's physical structure derived by \citet{jim12} is used in our calculations. We assume that all glycine within the mantles is released into the gas phase (abundance of $\sim$10$^{-8}$; Section$\,$\ref{hot}), and the line spectrum is corrected by the beam filling factor of the hot core within a single-dish telescope beam (in this case, the IRAM 30$\,$m telescope). The glycine lines at 1$\,$mm are brighter than in the pre-stellar core case. However, line confusion and line blending is a major issue in the detection of glycine in hot cores. Horizontal red lines are as in the upper panel.} 
\label{f2}
\end{center}
\end{figure*}

In our calculations, we also consider that glycine is FUV photo-desorbed together with water in L1544. Multilayer simulations of ice mantle formation indeed show that complex organics (e.g. methanol and formaldehyde) are formed in the outer layers of the mantle, while water is uniformingly distributed across these layers \citep{cup09,taq12}. Therefore, most of the complex organics are expected to be released alongside with water once the mantle outer layers are photo-desorbed\footnote{The penetrability of UV photons is in the range between 100$\,$nm and 200$\,$nm \citep[][]{mun13}.}. From this, the distribution of gas phase glycine is assumed to follow the water vapour abundance profile deduced in L1544 (see discussion in Section$\,$\ref{gly}), but scaled down by the fraction of solid glycine expected to be formed on ices. This can be calculated as: 

\begin{eqnarray}
\chi(Gly) (r) = \frac{\chi(Gly)_m}{\chi(H_2O)_m} \,\times\, \chi(H_2O) (r), \label{abun}
\end{eqnarray}

\noindent
where $\chi(Gly)_m$ is the abundance of solid glycine on ices, $\chi(H_2O)_m$ is the water abundance (relative to H$_2$) in the mantles \citep[$\chi(H_2O)_m$=7.25$\times$10$^{-5}$;][]{whi91}, and $\chi(H_2O)(r)$ is the distribution of gas phase water in L1544 \citep[][]{cas12}. We assume a solid glycine abundance of $\chi(Gly)_m$$\sim$10$^{-4}$ with respect to water on ices ($\sim$10$^{-8}$ with respect to H$_2$; see Section$\,$\ref{gly}). 

As shown in Figure$\,$\ref{f1}, the maximum gas-phase abundance of glycine reached in the model (red line) is $\chi(Gly)$$\sim$3$\times$10$^{-11}$, which corresponds to column densities $\sim$10$^{11}$$\,$cm$^{-2}$. This glycine abundance is reasonable since it is a factor of $\sim$100 lower than that of amino acetonitrile (NH$_2$CH$_2$CN, a precursor of glycine) measured in SgrB2(N) \citep[of $\sim$2$\times$10$^{-9}$;][]{bell08}. The abundance ratio $\frac{NH_2CH_2COOH}{NH_2CH_2CN}$$\sim$$\frac{1}{100}$ is of the same order of magnitude as the ratio between acetic acid and methyl cyanide in SgrB2(N) \citep[$\frac{CH_3COOH}{CH_3CN}$$\sim$$\frac{1}{200}$;][]{bell08}, as expected if the pairs glycine/amino-acetonitrile and acetic-acid/methyl-cyanide were formed by similar chemical pathways. 

The spectrum of glycine is obtained toward the core peak position considering spherical symmetry and LTE conditions. The 1D radiative transfer model follows the procedure described in \citet[][]{myers96} and \citet{devri05}, which was used for pre-stellar cores. The column density of glycine and the optical depth of every line are calculated as a function of radius and the radiative transfer is performed along the line-of-sight considering the core's velocity profile and linewidth of the emission at every position. The spectroscopic information of glycine is extracted from the CDMS \citep[][]{mue05}. 

Non-LTE calculations are not possible due to the lack of collisional coefficients of glycine with He or H$_2$. Assuming that these coefficients are of the same order as those derived for methyl formate \citep[HCOOCH$_3$, coefficients of $\sim$2-3$\times$10$^{-11}$$\,$cm$^3$$\,$s$^{-1}$;][]{fau14}, the estimated critical density for the glycine lines is $n_{cr}$$\sim$5-7$\times$10$^4$$\,$cm$^{-3}$, similar to the average H$_2$ density measured toward L1544 \citep[n(H$_2$)$\sim$7$\times$10$^4$$\,$cm$^{-3}$;][]{cas02b}. This implies that the glycine emission is probably not far from LTE. 

\subsection{Results}
\label{res}

In Figure$\,$\ref{f2} (upper panel), we show the simulated spectrum of glycine obtained between 60$\,$GHz and 250$\,$GHz considering the physical and kinematic structure of L1544 (Section$\,$\ref{mod}). While the lines at 1$\,$mm and 2$\,$mm show brightness temperatures $\leq$5$\,$mK, the 3$\,$mm and 4$\,$mm transitions (from 60$\,$GHz to 80$\,$GHz) have peak intensities $\geq$10$\,$mK. The brighter emission at 3$\,$mm and 4$\,$mm is due to the combination of low values of E$_{\rm u}$ ($\leq$30 K) and large upper level degeneracies, $g_{\rm u}$, for these lines. Since E$_{\rm u}$ progressively increases for higher frequencies, the higher energy levels are more difficult to populate making their expected line intensities very low. The glycine emission is optically thin across the core because the maximum optical depths attained for the brightest glycine lines are $\leq$10$^{-4}$.  
 
Since the glycine emission is extended across the core, this emission is expected to fill the beam of the telescope and the brightness temperature, T$_{\rm B}$, will be similar to the main beam temperature, T$_{\rm mb}$, measured by the telescope (T$_{\rm B}$$\sim$T$_{\rm mb}$, beam-filling factor $\sim$1). This has also been considered in previous studies of pre-stellar cores where radiative transfer calculations were compared to observations \citep{myers96,lee01}. Therefore, our simulations show that {\it for reasonable abundances of glycine in the gas-phase, the glycine lines could reach detectable levels in pre-stellar cores between 60$\,$GHz and 80$\,$GHz.}

We note that even if glycine did not follow the gas-phase water profile deduced for L1544 \citep{cas12}, peak intensities $\geq$10$\,$mK would still be measured between 60$\,$GHz and 80$\,$GHz if a constant glycine abundance of 1.5$\times$10$^{-11}$ were present in the gas phase in this core. This gas-phase abundance is the lowest level that could make glycine detectable in cold pre-stellar cores. 

\begin{deluxetable}{lccccc}
\tablecaption{Sample of brightest glycine lines predicted for L1544\label{tab1}}
\tablewidth{0pt}
\tablehead{
\colhead{Line} & Transition & \colhead{Frequency\tablenotemark{a}} & \colhead{$A_{\rm ul}$} & \colhead{$E_{\rm u}$} & \colhead{$g_{\rm u}$} \\
& & \colhead{(MHz)} & \colhead{(s$^{-1}$)} & \colhead{(K)} & }
\startdata

1 & 10$_{1,9}$$\rightarrow$9$_{1,8}$ & 67189.12 & 1.3$\times$10$^{-6}$ & 18.8 & 63 \\ 
2 & 10$_{3,8}$$\rightarrow$9$_{3,7}$ & 68323.70 & 1.3$\times$10$^{-6}$ & 21.1 & 63 \\ 
3 & 10$_{3,7}$$\rightarrow$9$_{3,6}$ & 71611.56 & 1.5$\times$10$^{-6}$ & 21.5 & 63 \\ 
4 & 11$_{2,10}$$\rightarrow$10$_{2,9}$ & 71646.39 & 1.6$\times$10$^{-6}$ & 22.4 & 69 \\ 
5 & 10$_{2,8}$$\rightarrow$9$_{2,7}$ & 71910.30 & 1.6$\times$10$^{-6}$ & 20.2 & 63 \\ 
6 & 12$_{1,12}$$\rightarrow$11$_{1,11}$ & 72559.35 & 1.7$\times$10$^{-6}$ & 23.3 & 75 \\ 
7 & 12$_{0,12}$$\rightarrow$11$_{0,11}$ & 72601.11 & 1.7$\times$10$^{-6}$ & 23.3 & 75 \\ 
8 & 11$_{1,10}$$\rightarrow$10$_{1,9}$ & 72841.25 & 1.7$\times$10$^{-6}$ & 22.3 & 69 \\ 
9 & 11$_{2,9}$$\rightarrow$10$_{2,8}$ & 78524.63 & 2.1$\times$10$^{-6}$ & 24.0 & 69 \\ 

\enddata

\tablenotetext{a}{Extracted from the CDMS \citep[][]{mue05}.}

\end{deluxetable}

\subsection{Formation and Destruction Mechanisms of Glycine}
\label{gly}

In its solid form, the amount of glycine that can be synthesized on ices has been investigated in laboratory experiments of ion- and UV-irradiated interstellar ice analogs. \citet{mun02} and \citet{ber02} reported the formation of glycine on UV-irradiated ices with fractions $\sim$0.004-0.025\% with respect to water. This implies solid glycine abundances of $\sim$4$\times$10$^{-9}$-2.5$\times$10$^{-8}$ with respect to H$_2$. As explained in \citet[][]{ber02}, the UV photon flux used in these experiments corresponds to the typical interstellar dose within dense molecular clouds for cloud ages $\sim$10$^5$$\,$yr and visual extinctions $A_v$$\sim$5$\,$mag. 

Experiments of ion-irradiation on interstellar ice analogs also synthesize glycine with column densities of $\sim$10$^{13}$$\,$cm$^{-2}$ \citep{holt05}. The interaction of energetic electrons in the track of a cosmic-ray impact on a dust grain yields HOCO and NH$_2$CH$_2$ neighbouring complexes which subsequently recombine forming glycine \citep{holt05}. The formation of glycine via the reaction HOCO + NH$_2$CH$_2$$\rightarrow$NH$_2$CH$_2$COOH requires no entrance barrier, and thus, it is feasible at 10$\,$K \citep[][]{holt05}. Therefore, even if glycine were not photo-desorbed from ices in L1544, column densities of $\sim$10$^{11}$$\,$cm$^{-2}$ could still be found in the gas phase assuming a 1\% efficiency for the desorption of glycine via cosmic ray impacts \citep[][]{leg85} and/or exothermic surface reactions \citep[][]{garr07}. These column densities are similar to those considered in our modelling (Section$\,$\ref{res}). 

The destruction rate of glycine by pure gas-phase reactions at low temperatures (mainly with H$_3$$^+$, H$_3$O$^+$ and HCO$^+$) is expected to decrease the abundance of this molecule by a factor of 10 in $\sim$2$\times$10$^5$$\,$yrs (as derived for methyl formate, a precursor of glycine, for which its gas phase reaction rates are relatively well-constrained). However, in the cold gas of pre-stellar cores, other processes such as surface diffusion by tunneling effects of atomic oxygen \citep{mini14}, cosmic rays induced diffusion \citep[or CRID;][]{reb14}, and episodic chemical explosions on ices \citep{raw13} could also enhance the abundance of COMs by factors from a few to $\sim$1000 in both the gas phase and on ices \citep{reb14}. In particular, CRID is extremely efficient at visual extinctions $A_v$$\sim$3-4$\,$mag, i.e. right at the extinction where water and glycine show their maximum gas-phase abundance in L1544 (Figure$\,$\ref{f1}). In these regions, COM formation on ices is predicted to occur at much higher rates than UV photo-destruction \citep{reb14}, suggesting that any possible UV photo-dissociation of glycine after desorption would be counteracted by CRID COM formation. We also note that if the formation of glycine in pre-stellar cores occurred via cosmic ray (ion) irradiation, photo-dissociation of glycine would not represent an issue. From all this, it seems reasonable to assume that glycine abundances as low as $\sim$10$^{-11}$ could be present in the cold gas of pre-stellar cores, as considered in our calculations.  

\section{Comparison with massive hot cores}
\label{hot}

For comparison, in Figure$\,$\ref{f2} (lower panel) we also show the LTE spectrum of glycine calculated for a hot molecular core from 60$\,$GHz to 250$\,$GHz. This is to illustrate the changes in the predicted spectrum of glycine at temperatures typical of hot cores. For our calculations, we have selected the AFGL2591 hot core for which its internal physical structure is well constrained \citep[][]{van99,jim12}. The hot core's physical structure, corrected by its new distance \citep[$\sim$3$\,$kpc;][]{rygl12}, is taken from \citet{jim12}. Since the gas/dust temperatures in the core lie above 130$\,$K, we assume that all glycine on ices is released into the gas phase [$\chi(Gly)$$\sim$10$^{-8}$]. We make predictions for a source size $\sim$1.5$"$ \citep[$\sim$4500$\,$AU;][]{jim12}, and for a single-dish telescope such as the IRAM 30$\,$m telescope\footnote{The derived T$_{\rm B}$ needs to be corrected by the beam filling factor defined as $\theta_s^2/[\theta_s^2+\theta_b^2]$, with $\theta_s$ the source size and $\theta_b$ the Gaussian beam size. $\theta_b$ is given by the diffraction limit \citep[][]{maret11}.}.  

Despite the larger abundance of glycine in hot sources, Figure$\,$\ref{f2} (lower panel) shows that the 3$\,$mm and 4$\,$mm glycine lines are factors $\sim$3-15 weaker in hot sources than in cold objects. This is due to: i) the lower efficiency to populate the low energy levels of glycine at temperatures $\geq$130$\,$K; and ii) the large beam dilution produced by the single-dish telescope. If observed with interferometers \citep[beam sizes $\sim$7-17$"$, as in][]{jon07}, the 3$\,$mm peak intensities would still lie below 50$\,$mK, which could partly explain why previous searches of glycine in hot cores with ATCA were not successful (upper limits to the column density of glycine $\leq$10$^{14}$-10$^{15}$$\,$cm$^{-2}$). As noted in Section$\,$\ref{intro}, the presence of more abundant COMs in hot cores whose emission has linewidths of some km$\,$s$^{-1}$, is a major issue in the detection/identification of glycine in hot sources due to line blending and line confusion.

\section{Implications for ALMA Band 2}
\label{alma}

The advent of new instrumentation providing higher-sensitivity observations such as the Atacama Large Millimeter Array (ALMA), opens up the possibility to detect a large number of COMs of biochemical interest such as glycine in low-mass star forming regions. Our modelling of Section$\,$\ref{pre-stellar} is based on reasonable assumptions of the dominant chemical processes in pre-stellar cores and of the glycine abundance on ices and in the gas phase. This modelling not only shows that glycine could be detected in these cold objects if abundances as low as $\sim$10$^{-11}$ were present in the gas phase, but more importantly, that this emission, if present, should be searched for in the frequency range between 60$\,$GHz and 80$\,$GHz where the predicted glycine intensities are $\geq$8-10$\,$mK (Figure$\,$\ref{f2}). This is the frequency range that the Band 2 receivers of ALMA will cover (from 67$\,$GHz to 90$\,$GHz)\footnote{See http://www.almaobservatory.org/en/about-alma/how-does-alma-work/technology/front-end. The frequency range 60-67$\,$GHz cannot be observed from the ground because of the high atmospheric opacity produced by ozone.}, making this band very well suited for the discovery of glycine and other pre-biotic molecules in Solar-type system precursors. In Table$\,$\ref{tab1}, we summarize the glycine transitions that represent the best targets to detect glycine in cold sources. We note that other glycine lines with E$_{\rm u}$$\leq$20$\,$K do exist at frequencies $\leq$60$\,$GHz. However, these transitions have $A_{\rm ul}$ coefficients $\leq$10$^{-6}$$\,$s$^{-1}$ \citep[CDMS;][]{mue05}, making their detection in cold objects more difficult. All this suggests that the discovery of glycine (and other pre-biotic species) could be possible toward the coldest phases in the formation of Solar-type systems with near-future instrumentation.

\acknowledgments

We thank Guillermo Mu\~noz-Caro for his insight into the laboratory experiments producing glycine on ices. We also acknowledge an anonymous referee whose comments helped to improve the original version of the manuscript. The research leading to these results has received funding from the People Programme (Marie Curie Actions) of the European Union's Seventh Framework Programme (FP7/2007-2013) under REA grant agreement PIIF-GA-2011-301538. P.C. acknowledges the financial support of successive rolling grants awarded by the UK Science and Technology Funding Council and of the European Research Council (ERC; project PALs 320620). S.V. acknowledges financial support from the (European Community's) Seventh Framework Program [FP7/2007–2013] under grant agreement 238258.



\begin{thebibliography}{}

\bibitem[Bacmann et al.(2012)]{bac12}
Bacmann, A., Taquet, V., Faure, A., Kahane, C., \& Ceccarelli, C. 2012, \aap, 541, L12

\bibitem[Belloche et al.(2013)]{bell13}
Belloche, A., M\"uller, H. S. P., Menten, K. M., Schilke, P., \& Comito, C. 2013, \aap, 559A, 47B

\bibitem[Belloche et al.(2008)]{bell08}
Belloche, A., et al. 2008, \aap, 482, 179

\bibitem[Bernstein et al.(2002)]{ber02}
Bernstein, M. P., Dworkin, J. P., Sandford, S. A., Cooper, G. W., \& Allamandola, L. J. 2002, Nature, 416, 401

\bibitem[Caselli et al.(1999)]{cas99}
Caselli, P., Walmsley, C. M., Tafalla, M., Dore, L., \& Myers, P. C. 1999, \apj, 523, L165

\bibitem[Caselli et al.(2002a)]{cas02a}
Caselli, P., Walmsley, C. M., Zucconi, A., Tafalla, M., Dore, L., \& Myers, P. C. 2002, \apj, 565, 331

\bibitem[Caselli et al.(2002b)]{cas02b}
Caselli, P., Benson, P. J., Myers, P. C., \& Tafalla, M. 2002, \apj, 572, 238

\bibitem[Caselli et al.(2012)]{cas12}
Caselli, P., et al. 2012, \apj, 759, L37 

\bibitem[Crapsi et al.(2005)]{cra05}
Crapsi, A., Caselli, P., Walmsley, C. M., Myers, P. C., Tafalla, M., Lee, C. W., \& Bourke, T. L. 2005, \apj, 619, 379

\bibitem[Crapsi et al.(2007)]{cra07} Crapsi, A., Caselli, P., Walmsley, M. C., \& Tafalla, M. 2007, \aap, 470, 221

\bibitem[Cunningham et al.(2007)]{cun07}
Cunningham, M. R., et al. 2007, \mnras, 376, 1201

\bibitem[Cuppen et al.(2009)]{cup09}
Cuppen, H. M., van Dishoeck, E. F., Herbst, E., \& Tielens, A. G. G. M. 2009, \aap, 508, 275

\bibitem[de Vries \& Myers(2005)]{devri05}
de Vries, C. H., \& Myers, P. C. 2005, \apj, 620, 800

\bibitem[Ehrenfreund et al.(2001)]{ehr01}
Ehrenfreund, P., Glavin, D. P., Botta, O., Cooper, G., \& Bada, J. L. 2001, PNAS, 98, 2138

\bibitem[Ehrenfreund \& Charnley(2000)]{ehr00}
Ehrenfreund, P., \& Charnley, S. B. 2000, ARA\&A, 38, 427

\bibitem[Elias(1978)]{elias78}
Elias, J. H. 1978, \apj, 224, 857	

\bibitem[Elsila et al.(2009)]{els09}
Elsila, J. E., Glavin, D. P., \& Dworkin, J. P. 2009, M\&PS, 44, 1323

\bibitem[Faure et al.(2014)]{fau14}
Faure, A., Remijan, A. J., Szalewicz, K., \& Wiesenfeld, L. 2014, accepted, arXiv1401.1136F

\bibitem[Garrod(2013)]{garr13}
Garrod, R. T. 2013, \apj, 765, 60

\bibitem[Garrod et al.(2007)]{garr07}
Garrod, R. T., Wakelam, V., \& Herbst, E. 2007, A\&A, 467, 1103G

\bibitem[Garay \& Lizano(1999)]{gar99}
Garay, G., \& Lizano, S. 1999, \pasp, 111, 1049

\bibitem[Glavin et al.(2006)]{gla06}
Glavin, D. P., Dworkin, J. P., Aubrey, A., Botta, O., Doty, J. H., Martins, Z., Bada, J. L. 2006, M\&PS, 41, 889

\bibitem[Hasegawa \& Herbst(1993)]{has93}
Hasegawa, T. I., \& Herbst, E. 1993, \mnras, 261, 83

\bibitem[Hollenbach et al.(2009)]{hol09}
Hollenbach, D., Kaufman, M. J., Bergin, E. A., \& Melnick, G. J. 2009, \apj, 690, 1497

\bibitem[Holtom et al.(2005)]{holt05}
Holtom, P. D., Bennett, C. J., Osamura, Y., Mason, N. J., \& Kaiser, R. I. 2005, \apj, 626, 940

\bibitem[Jim\'enez-Serra et al.(2012)]{jim12}
Jim\'enez-Serra, I., Zhang, Q., Viti, S., Mart\'{\i}n-Pintado, J., \& de Wit, W.-J. 2012, \apj, 753, 34

\bibitem[Jones et al.(2007)]{jon07}
Jones, P. A., Cunningham, M. R., Godfrey, P. D., \& Cragg, D. M. 2007, \mnras, 374, 579

\bibitem[Keto \& Caselli(2010)]{ket10}
Keto, E., \& Caselli, P. 2010, \mnras, 402, 1625

\bibitem[Keto et al.(2014)]{ket14}
Keto, E., Rawlings, J. C., \& Caselli, P. 2014, \mnras, submitted

\bibitem[Kuan et al.(2003)]{kuan03}
Kuan, Y.-J., Charnley, S. B., Huang, H.-C., Tseng, W.-L., \& Kisiel, Z. 2003, \apj, 593, 848

\bibitem[Lee et al.(2001)]{lee01} 
Lee, C. W., Myers, P. C., \& Tafalla, M. 2001, ApJS, 136, 703

\bibitem[L\'eger et al.(1985)]{leg85}
L\'eger, A., Jura, M., \& Omont, A. 1985, A\&A, 144, 147L

\bibitem[Marcelino et al.(2007)]{mar07}
Marcelino, N., et al. 2007, \apj, 665, L127

\bibitem[Maret et al.(2011)]{maret11}
Maret, S., Hily-Blant, P., Pety, J., Bardeau, S., \& Reynier, E. 2011, A\&A, 526, 47

\bibitem[Minissale et al.(2014)]{mini14}
Minissale, M., Congiu, E., Dulieu, F. 2014, {\it Journal of Chem. Phys.}, 140, 074705

\bibitem[M\"uller et al.(2005)]{mue05}
M\"uller, H. S. P., Schl\"oder, F., Stutzki, J., \& Winnewisser, G. 2005, J. Mol. Struct. 742, 215

\bibitem[Mu\~noz Caro et al.(2002)]{mun02}
Mu\~noz Caro, G. M., et al. 2002, \nat, 416, 403

\bibitem[Mu\~noz Caro \& Dartois(2013)]{mun13}
Mu\~noz Caro, G. M., \& Dartois, E. 2013, Chem. Soc. Rev., 42, 2173

\bibitem[Myers et al.(1996)]{myers96}
Myers, P. C., Mardones, D., Tafalla, M., Williams, J. P., \& Wilner, D. J. 1996, \apj, 465, L133

\bibitem[Pagani et al.(2007)]{pag07} Pagani, L., Bacmann, A., Cabrit, S., \& Vastel, C. 2007, \aap, 467, 179

\bibitem[Pizzarello et al.(1991)]{piz91}
Pizzarello, S., Krishnamurthy, R. V., Epstein, S., \& Cronin, J. R. 1991, GeCoA, 55, 905

\bibitem[Rawlings et al.(2013)]{raw13}
Rawlings, J. M. C., Williams, D. A., Viti, S., \& Cecchi-Pestellini, C. 2013, \mnras, 430, 264

\bibitem[Reboussin et al.(2014)]{reb14}
Reboussin, L., Wakelam, V., Guilloteau, S., \& Hersant, F. 2014, MNRAS, submitted (arXiv:1403.5189R)

\bibitem[Rygl et al.(2012)]{rygl12}
Rygl, K. L. J., et al. 2012, \aap, 539A, 79R

\bibitem[Snyder et al.(2005)]{sny05}
Snyder, L. E., et al. 2005, \apj, 619, 914

\bibitem[Taquet et al.(2012)]{taq12}
Taquet, V., Ceccarelli, C., \& Kahane, C. 2012, \aap, 538A, 42T

\bibitem[van der Tak et al.(1999)]{van99}
van der Tak, F. F. S., van Dishoeck, E. F., Evans, N. J., II, Bakker, E. J., \& Blake, G. A. 1999, \apj, 522, 991

\bibitem[van Dishoeck \& Blake(1998)]{van98}
van Dishoeck, E. F., \& Blake, G. A. 1998, ARA\&A, 36, 317

\bibitem[Ward-Thompson et al.(1999)]{ward99}
Ward-Thompson, D., Motte, F., \& Andre, P. 1999, \mnras, 305, 143

\bibitem[Whittet \& Duley(1991)]{whi91}
Whittet, D. C. B., \& Duley, W. W. 1991, A\&ARv, 2, 167

\end{thebibliography}
\end{document}